Chapter 4

Crowds for Clouds: Recent Trends in Humanities Research Infrastructures

Tobias Blanke, ConnyKristel, Laurent Romary

**Introduction**

The humanities community has been remarkably successful in applying for research infrastructure funding in Europe. For instance, in the first call for the Horizon 2020 programme on Integrating Activities 'to open up key national and regional research infrastructures to all European researchers, (…)' (Commission, 2014), from over 60 proposals 10 were finally chosen for the first round of funding. Out of these selected proposals, two proposals came directly from humanities communities, which makes them one of the most successful communities in this call. Humanities seemed tohave convincingly argued that they need transnational research opportunities and through the digital transformation of their disciplines also have the means to proceed with it on an up to now unknown scale. The Roman Empire cannot be studied with resources in Italy alone but we need to incorporate the many modern countries that were part of that empire.

The digital transformation of research and its resources means that many of the artifacts, documents, materials, etc. that interest humanities research can now be combined in new and innovative ways. A recent project in the UK links documents about court cases in the Old Bailey system with the fate of deportees in Australia (http://www.digitalpanopticon.org/). Such projects are emerging because there is a need in the corresponding communities to expand their research environments. Digital transformations have brought about new actors and practices in all areas of researching culture. More and more cultural objects are integrated into the digital space through processes of datafication(Kitchin, 2014, Mayer-Schönberger and Cukier, 2013), while infrastructures working with these digital objects provide a sense of stability and continuity (Edwards et al., 2009).

At the same time, digital transformations offer new possibilities for humanities research to reassemble new socio-technical methods and devices (Ruppert et al., 2013) in order to explore society and culture. Due to the digital transformations, (big) data and information have become central to the study of culture and society. Big data is not limited to the sciences and large-scale enterprises. With more than 7 billion people worldwide, large amounts of data are produced in social and cultural interactions, while we can look back onto several thousand years of human history that have produced vast amounts of cultural records.

Humanities research infrastructures manage, organise and distribute this kind of information and many more data objects as they becomes relevant for social and cultural research. Edwards has explored infrastructures as global sociotechnical systems and as characteristics of modern society, where one lives within and by means of infrastructures (Edwards et al., 2009). Research infrastructures, in particular, helped disciplines to redefine themselves around a shared set of devices that support their research. Humanities research infrastructures have been theorized as digital ecosystems without a centre and constituted through heavily interconnected online platforms (Anderson and Blanke, 2012).

Along these lines, the European Commission defines research infrastructures as 'facilities, resources or services of a unique nature that have been identified by research communities to conduct top level activities in their fields. They may be single sited, distributed or virtual.' (ESFRI, 2010) They 'often produce large amounts of data requiring data management'. In the case of humanities research infrastructures much of the 'data' for integration are not a product of the infrastructure itself but are the primary source materials, produced as a result of the activities of cultural heritage institutions; mostly in archives and libraries. Large-scale digitization efforts have recently begun to create digital surrogates for human history. Especially, the European Union seems committed to digitize and present its cultural heritage online. As of end of 2014, its Cultural Heritage aggregator Europeana has made available over 30m digital objects through its portal (Bernipe, 2014).

In this chapter, we concentrate on recent trends in humanities research infrastructures. We observe the common practices that have emerged in various large-scale transnationally operating infrastructure projects. We focus on research infrastructures rather than (digital) library and archive integration projects such as Europeana, because research infrastructures share the overall final aim to action research. Europeana on the other hand aims to primarily fulfill the needs of a culturally interested public rather than research community.

Hidden in this broad distinction are of course many commonalities between these initiatives. They all make use of a similar set of technologies for integrating digital collections, while the primary user community of libraries and in particular archives have been humanities researchers. Finally, they are concerned with access to and long-term availability of cultural collections. Nevertheless, the differences between the initiatives are defined enough to merit a focus on research infrastructures, as a distinct undertaking in the humanities.

**Humanities Research Infrastructures**

The link between humanities and their infrastructures has been strong in the past though sometimes hidden. Some like the British historian Marina Warner have even defined humanities themselves 'as infrastructure' for critical thinking about culture and how 'people connect' (Preston, 2015). Furthermore, literary scholars and philosophers are regulars in research libraries, while historians lead many archives.

The UK's Times Higher Education summarizes the view that the humanities too need firm foundations and a new culture of sharing, both of which are signature elements of research infrastructures (Reisz, 2014). There have been several attempts to define what should interest the humanities with regards to infrastructures. In order to discriminate humanities infrastructures from scientific ones, the digital humanities scholar Svensson famously asked for a 'conceptual cyberinfrastructure' that can be 'seen as a set of underlying ideas that provide the ideational grounding of a particular instance of research infrastructure.' (Svensson, 2011). Svensson wants a humanities infrastructure to be first and foremost about humanities, while he clearly sees an interest from those in the sciences in humanities research infrastructures: 'Dan Atkins, then head of the US NSF Office of Cyberinfrastructure, [demanded] that the humanities and social sciences step up and show leadership in relation to the issue of future cyberinfrastructure'. (Svensson, 2011). While in the US the humanities have not managed to sustain larger projects of integrated infrastructures, the Europeans have seen various successful projects.

A couple of years ago, Manfred Thaller from Cologne brought together a number of digital humanities researchers for a seminar on the current state of affairs in the digital humanities (Thaller, 2012). The controversial discussions included an exchange on the nature of research infrastructures and the relationship of digital humanities to them. The Dutch Digital Humanities scholarJoris van Zundert asked 'If you build it, will we come? Large scale digital infrastructures as a dead end for digital humanities.' (Van Zundert, 2012). He held up his own experience in the international network InterEdition against three examples of 'big all-encompassing all-serving digital infrastructures', which he considered meaningless for the development of humanities. He cited Bamboo in the US as well as the European initiatives DARIAH, which brings together digital humanities initiatives in Europe, and CLARIN, which develops language resources.

Bamboo's funding has in the meantime run out (Dombrowski, 2014). We thus focus on DARIAH (http://www.dariah.eu) and CLARIN (http://www.clarin.eu/) and can add their related more domain-specific infrastructures such as ARIADNE (http://www.ariadne-infrastructure.eu/), which focuses on archaeology; CENDARI (http://www.cendari.eu/) working on medieval and First World War resources; EHRI (www.ehri-project.eu), concerned with Holocaust research; and finally IPERION (http:// www.iperionch.eu), which is about material research on cultural heritage. The latter four projects are all associated with DARIAH, which has through them become a platform for transnational humanities collaborations, while CLARIN has continued to have a transforming impact on language research in Europe.

While van Zundert's criticism was not based on the just cited domain infrastructures, it could have also applied to them. ARIADNE is following the assumption that archaeologists have a need to integrate their research data, while van Zundert seems to imply (Van Zundert, 2012) that digitized humanities resources are available through libraries and in particular their own large-scale collaborations like Europeana. EHRI, on the other hand, was focused in its first phase from 2010-15 on integrating archival material on the Holocaust from relevant sources across Europe. Thus, the project made a strong effort to integrate three communities: history, digital humanities and archives. The direct work with archivists in Holocaust institutions was considered especially important, because most of the material on Holocaust is not digitally available yet. The Dutch NIOD, who is also the coordinator of EHRI and a well-funded Dutch national archive on Holocaust and genocides, has archives of approx. 2.5 kilometres long and thus without doubt in our world pretty big data. Only 2% of it, however, is even available in a digital format though not always accessible online. The objective for 2016 is 7%. Any kind of research infrastructure project based on data that aims to link across archives needs to therefore find ways of joining up the analogue and digital information.

The final two domain infrastructures IPERION and CENDARI concentrate on what van Zundert criticizes as 'snowballing IT-based methodological innovation into a humanities domain' (Van Zundert, 2012). IPERION assembles access to humanities research to support the preservation of material cultural heritage. The use of IT-based instruments to produce 3D visualizations of artworks and other heritage objects has enhanced the capacities of curators. IPERION brings their knowledge together. CENDARI has the unusual task by the Commission to develop tools for medieval and first world war historians and has demonstrated that, while these historical communities are distinct, they also share needs in a methodological commons (Anderson et al., 2010) of researching archives.

Van Zundert's main concern, however, are of course CLARIN and DARIAH and here especially CLARIN, which was further advanced at the time of the workshop in Cologne. He states: 'Being standards-driven, institutionally bound, and at worst enforcing specific implementations, they are platforms of exclusiveness.' (Van Zundert, 2012). Against them, he demands 'open platforms' based on 'agile processes'. In a contribution to the same workshop, Anderson and Blanke (2012) gave this idea of open platforms based on agile processes the name 'digital ecosystems'. The concept of ecosystems was belittled in the past but is more accepted now in our experience and widely used to describe a loosely coupled organization of services and activities.

Blanke (2014)analyses how the concept of digital ecosystems summarizes decentralized digital work; precisely because of its contested origin in biological sciences. According to Briscoe and Sadedin (2009), a natural environment consists of ecosystems, which in turn are inhabited by habitats and communities. It is easy to form an analogy here so that populations are crowds forming on the Internet, or the collaboration of large numbers of humans on a common task. The habitats are the platforms or the 'clouds' crowds work on. Together, communities and habitats build niches or, in our sense, applications and services that are built around them. Mark Zuckerberg(Blanke, 2014) has pinned his hope on digital ecosystems, because the smaller Facebook can only compete with the giants of Google and Microsoft if it manages to organize an ecosystem, with which it can effectively integrate outside innovation.

According to the IEEE Digital Ecosystem conference, digital ecosystems are 'loosely coupled, domain-specific [...] communities which offer cost-effective digital services and value- creating activities' (IEEE Digital Ecosystem, 2007). In this definition, digital ecosystems are derived from communities, or crowds with a set purpose, rather than technologies, just like in the Facebook case, where Mark Zuckerberg(Blanke, 2014) first associates developers and users with digital ecosystems, because in the digital ecosystem they define their services and take control. Therefore, the technologies of the digital ecosystem need to be thought of from the perspective of the crowds and do not define what crowds are. This requires a commitment to an open platform, which can only be developed using open standards.

To think of research infrastructures as digital ecosystems, thus entails the commitment to identifying them as services that are built around communities. Infrastructures are then the sum and integration of these services that are shared through a platform. Communities as crowds work together on a common goal and become the most important resource for the sustainability of the infrastructure. Especially in the arts and humanities, characterized by high diversity as well as limited overall funding opportunities, no single community alone can achieve an investment over a sustained period of time. Software is commonly co-developed with open-source licenses, while code is shared so that any improvement goes directly back into the software. The knowledge in existing software is reverse-engineered and made future-proof.

**European Humanities Research Infrastructures as Digital Ecosystems**

In this section, we will investigate how for the above-cited six European larger humanities research infrastructures a digital ecosystem is emerging as the collaboration between crowds and clouds. We begin with CLARIN, which is concentrated on the particular community using language resources. Many of its partners have worked together for decades. CLARIN is organized around Centres, which provide services to each other (https://www.clarin.eu/clarin-eric-datatables/centres). These are mature and certified through a formal procedure, registered in a central database (https://centres.clarin.eu/) and compliant with a range of requirements in order to be allowed into the CLARIN network.

With its Centre structure, CLARIN is the most formally structured infrastructure of the ones discussed here. It has managed to bring together many parts of the European community that work with language resources. CLARIN's key paper (Váradi et al.) has mainly been cited by the language resources community according to Google Scholar.[1] CLARIN delivers a central repository of language resources, a cloud so to say, together with basic technologies to manage this cloud. This way, CLARIN fulfills one of van Zundert's central demands: 'We all have email, why not let us all have access to an academic computing cloud?' (Van Zundert, 2012: 18). CALRIN's strength is clearly its concentration on a cloud of language resources that serve researcher crowds with specific needs. Its

If CLARIN's strength is its depth, then DARIAH's is its breadth(Blanke et al., 2011). DARIAH's main objective is to organize national initiatives in Digital Humanities in its various member states, which are all more or less independent. DARIAH's second aim is to develop a platform (cloud) for various other trans-national European digital arts and humanities initiatives, some of which we cited above and all of which are independent entities. Its central form of organization is the working group (https://dariah.eu/library/resources.html), where researchers come together to collaboratively develop activities and services. The working group model is clearly copied from successful Web 2.0 architectures that often succeed in realizing a collective platform by working together in special interest communities (Blanke et al., 2011).

---

[1] Google Scholar search 14/10/15:
https://scholar.google.co.uk/scholar?cites=9903500876139790736&as_sdt=2005&sciodt=0,5&hl=en

DARIAH working groups (WG) are based on a three-step approach:

1. **Conceptualisation:** Members of DARIAH develop a concept for the WG, aggregate participants and means as well as define an action plan. The primary outcome is a proposal to the DARIAH community that meets the WG requirements.
2. **Implementation:** At this stage, efforts are integrated and the connection to further communities outside DARIAH is established.
3. **Service:** Finally, the service is enabled with guaranteed hosting and sustainable funding through DARIAH. The service is published to the community.

There are no specific criteria for working groups but they need to be collective efforts and need to comply with the DARIAH mission (http://dariah.eu/about/mission.html). Current working groups include visual media, education, a Digital Humanities course registry, digital annotations, service registries, a digital methods observatory, etc.

DARIAH's organization is the most bottom-up we could find. While CLARIN's strength is its organization in many stable centres, DARIAH is organized around distinct collaborations of its crowds. This makes the construction less stable but allows for a stronger dynamic and better flexibility to adjust to new research areas. CLARIN and DARIAH can implement these structures, as they are stable membership organizations within the European Strategy Forum on Research Infrastructures framework (ESFRI, 2010). The rest of the observed initiatives are in comparison traditional projects, funded mainly and exclusively by the European Commission. Nevertheless, these projects have shown remarkable innovations of how services and their platforms can be developed around communities.

CENDARI and EHRI both work with communities that are engaged in archival research. Both assemble historians, archivists and specialists in the Digital Humanities to make archival documents, which were often collected under difficult circumstances, available and accessible for research and to develop innovative techniques and methodologies for analysis and interpretation of such documents. Both projects concentrate in particular on those archives that are not part of larger infrastructures and/or are 'hidden' to most researchers. CENDARI has a stronger technical focus with more work dedicated to developing Digital Humanities solutions (Gartner and Hedges, 2013), while EHRI concentrates mainly on archival integration (Speck et al., 2014). CENDARI collaborates with large clusters of special collections in The European Library. EHRI's focus is on individual archives. Both, however, share the commitment that exploration of archives in the digital age is only possible if historians form collaboration crowds with archivists, which also include digital experts. This way, the archive becomes a new research space for historical discovery. CENDARI's corresponding aim is thus a discovery space, which can be considered as its cloud (http://www.cendari.eu/description-of-work/).

Both projects develop a virtual observatory for research collections as a key component of their work. EHRI has already released its 'virtual observatory' for Holocaust collections (Blanke and Kristel, 2013), which is a cloud of collection descriptions. The observatory is based on an integration of descriptions from partner sites and new descriptions arising from EHRI's own investigation work. The main objective was to allow researchers to retrieve information about archival sources that pertain to a particular research theme across repositories such as 'find all information about the departure of Dutch prisoners to the Terezín Ghetto'.

Through the EHRI Virtual Observatory (https://portal.ehri-project.eu/), researchers are now able to query disparate archives. To this end a wide range of different types of institutions holding relevant archival material were identified including national and regional archives, memory institutions, museums, and local and private collection holders. Each type of archive poses a unique challenge in terms of integrating their material, harmonizing the metadata and publishing it in an integrated portal. As new collections and other material are continuously discovered, the identification and integration work needs to continue all the time. EHRI thereby also sets an example for other domains engaged in historical research, showing how 'big data' on human history can be developed. Both EHRI and CENDARI thus develop a research environment that will allow their respective communities to cope with the big research data they are faced with.

The technical design challenge (Blanke et al., 2013) for archival research infrastructures such as EHRI and CENDARI is to innovate a dynamic, research-driven collection cloud, where new material is permanently discovered, added and analyzed. Their work has to rethink some of our assumptions that stemmed from traditional work with relatively stable cultural heritage collections and develop an environment that is technically flexible enough to allow for the integration of heterogeneous material and that is social enough to allow researcher crowds to discover and analyze their material and make new connections.

IPERION works with curators and other researchers concerned with the material aspects of heritage, while ARIADNE develops devices for archaeologists. Both share their commitment to developing data infrastructures that go beyond CENDARI's and EHRI's connection to archives and concentrate on the full sharing of research data sets. Both projects are committed to long-term preservation and dissemination. Contrary to earlier cited community concerns about standards as limiting research freedom, for ARIADNE standards enable the free exchange between crowds. For ARIADNE, '[t]he main processes for making data understandable and shareable are standardization and registration.' (Aloia et al., 2014). Standards have to be part of any research data cloud. The standardization and sharing of data sets cannot simply be handed over from the communities to professional providers such as libraries. While they need to work together, for ARIADNE there remains a distinct need for subject-based repositories. As Julian Richards, one of the leaders of the ARIADNE project, has put it, few of the libraries are currently equipped with the knowledge or the platform to deal with complex data sets that incorporate many discipline-specific assumptions. 'Several studies have recognised the value of discipline-based repositories in developing stakeholder communities, avoiding fragmentation, and establishing discipline-specific data preservation expertise.' (Richards, 2012).

Working with complex data sets that are developed by highly diversified research projects to investigate cultural objects, poses particular challenges to the 'conceptual' integration of data in clouds. For researchers, the grand promises and challenges with regard to the integration of data sets lie in their 'linkability'. This is independent whether the data is collected in small independent projects, using unorganized spreadsheets and word documents, or in large joint projects such as IPERION and ARIADNE. All are research databases and contain interpretations and expressions of uncertainty. We simply do not know enough from the past to exactly determine when Queen X died or Volcano Y erupted. These datasets are part of what Buneman from the Digital Curation Centre has called 'curated databases' (Buneman et al., 2008), as they are human created and therefore full of inconsistencies.

ARIADNE works on a Linked Data cloud that conceptually integrates archaeological curated databases (Aloia et al., 2014), which will realize the promises of linkability. A number of earlier experiments at King's College London with integrating traditional humanistic data sets (Blanke et al., 2012) tried to find out whether with a Linked Data cloud, uncertainties can be reduced by linking the information in one data source with the information in another. The results were mixed. While the production of links is already an issue because of the heterogeneous nature of curated databases, it can be even more difficult to define how to sensibly consume information that is highly interlinked. Links are good but also confusing for human researchers. A possible solution will be a community effort like ARIADNE that takes into account the production as well as consumption needs effort.

In this section, we have discussed how all the humanities infrastructures attempt to develop service and activities clouds around communities. For CLARIN, this has meant to concentrate on dedicated centres as a core stable form of organization. DARIAH is an umbrella organization that develops its platform collectively in working groups. The four other projects we discussed share a concern for the specifics of humanities data whether it is held externally from the communities in archives or is produced directly by the community.

Those concentrated on archives are mainly concerned with how to translate the cultural heritage institutions' holdings into something that fits into the needs of digital research. Those projects concerned with data sets that are created by researchers themselves (or at least for them) need to overcome the limitations of research data sets, as they are common not just to the humanities but many other research disciplines. For humanities, this means a focus on concepts and Linked Data clouds as supporting integration and overcoming heterogeneity that are the result of how these data sets are produced and curated by research crowds.

**Enhancing and strengthening the crowd**

All our discussed examples demonstrate how important it is for humanities infrastructures to create crowds with their data. As computers fail with the complexity of knowledge in curated databases of culture, crowds become more important. Research infrastructures need to connect human brains to perform complex reasoning on data. 'Networks connect people as well as devices, and when they are cheap and easy to use it means that those intellectual tasks more efficiently performed elsewhere by other people can be broken out and distributed.' (Zittrain, 2008).

Following the idea of ecosystems, crowds stand next to clouds as equal components of an infrastructure (Blanke, 2014). This idea goes back to an early conceptualization of Amazon to have its crowds perform tasks where its clouds cannot help. It set up its Mechanical Turk system to achieve this. Taking the Amazon infrastructure view (Blanke, 2014), crowds collaborate in their Mechanical Turk platform as part of an emerging larger infrastructure that supports the new kinds of production and consumption of digital value. Amazon has chosen to offer its crowd-sourcing functionalities through the same interface by which its other services are accessible. The substitution of computer intelligence by human intelligence is hidden from the outside world. If the crowds work smoothly, the service seems as seamless as a computer service. 'Hidden away under the appearance of computer-generated work, crowds have been increasingly rendered visible only through the design of new infrastructures.' (Aradau and Blanke, 2013: 38)

Of course, in the Amazon view and its Mechanical Turk implementation, crowds are paid to contribute for often not very satisfying tasks. Amazon's crowd is not based on volunteer contributions, but each participant gets paid a small amount per each task completed. In Amazon's terminology, these are Human Intelligence Tasks (HITs), and requesters define tasks and upload data, while workers (aka Turkers, which is the name Amazon Mechanical Turk workers are known under) do tasks and get paid. Typical tasks include the identification of email addresses in texts or the labelling of images. Workers' rights do not seem to follow the Turkers. There is, for instance, no guarantee of payment after the job and there are no benefits. http://turkernation.com has a hall of shame of worst jobs for Turkers.

For non-commercial research infrastructures' interest in crowds, financial benefits cannot be an option. Malone et al. (2009) have mapped the 'genome of collective intelligence' from crowds. They define collective intelligence as 'groups of individuals doing things collectively that seem intelligent' (Malone et al., 2009: 2) and give three reasons why people would like to collaborate to appear intelligent: money; love or enjoyment of an activity; and glory when recognition is achieved among peers. All these are opposed to a hierarchical distribution of labour. In this section, we will investigate what the future of research infrastructure development in the humanities might learn from these crowd-based collective intelligence initiatives.

Humanities need to recognize that its infrastructure needs are specific and realize that they need to take human computing innovations seriously, as long as computers are as they are and as long as the funding for collections in the humanities will always be much less than in the sciences. As long as humanities collections are too complex and analogue for computers to effectively deal with, much more focussed work on crowds is needed, which are according to the ecosystem model a recognized form of infrastructure. In the past crowds and humanities were concentrated on enriching existing heritage collections (Holley, 2010), which can be a successful model to help cash-stripped heritage and humanities communities to add value to their data. However, there are other innovations in the larger crowd economy that we should recognize as particularly suited for addressing humanities computing challenges.

We have already discussed earlier that such challenges generally include a focus on working with data. Because in the humanities, this data comes in many different formats, has complex semantic relationships and generally does not comply with needs of algorithms, there is detailed work involved in preparing the data. OCRing, for instance, works nowadays well with standard print collections of newspapers but continues to struggle with hand-written documents as they are common to many historical recordings. Therefore many heritage institutions have begun to set up large-scale crowd-sourcing projects to transcribe their collections(Holley, 2010). One of the most successful heritage crowd-sourcing projects was the digitization of the Australian newspaper archives and is at the same time a good example of these kind of projects in the humanities. In 2007, the National Library of Australia began to digitize out-of-copyright newspapers. It used crowds to help correct OCR (optical character recognition) mistakes, and the public followed in large numbers and analyzed millions of lines of text (Holley, 2010). The Digital Humanities have taken up efforts to digitize collections with some standout projects of their own such as Transcribe Bentham (http://blogs.ucl.ac.uk/transcribe-bentham/) or Diaries of First World War soldiers (http://www.operationwardiary.org/).

However, we believe we can learn about the infrastructure future in the humanities from other crowd-sourcing innovations as well, because of the specific type of computation Digital Humanities projects entail. We have just discussed the complexities of humanities data that require human intelligence tasks rather than computational analysis. Another typical characteristics of Digital Humanities are, that humanities computation is generally speaking quite small and well contained. The tasks involved are often similar to each other such as setting up a website but not similar enough for computers to completely take over. Finally and most importantly, individual organizations in the Digital Humanities are not big enough to have the in-house resources that would enable them to exploit cultural resources sufficiently.

CrowdSPRING (http://www.crowdSPRING.com) is a typical example of an online marketplace, designed to address the kinds of challenges just discussed. It offers 'creative services' such as logo design, website development, etc. The clientele is anybody from large and small businesses to private people. They chose from the offerings of 'creatives' from over 200 countries, who present examples of their work. The model has been criticized as undermining traditional design work (Hyde, 2008) but has proven to be effective in distributing small well-contained creative tasks. 'Because buyers on crowdSPRING select from actual designs, designers on crowdSPRING submit work on spec. "Spec" is a short name for doing any work on a speculative basis, (…).' (Basecamp, 2008). The advantage for designers is that crowdSPRING offers project management services and militates against payment and legal risks.

Online marketplaces are booming (Laumeister, 2014) by offering effective management and allocation of small tasks for work that can only be done by humans. They would therefore be perfectly suited models for humanities computing tasks. Blanke et al. (2011)develop a model for such marketplaces in DARIAH and demonstrate that the challenges involved are social. The marketplace platform has no value to its users unless both those who need to define tasks as well as those willing to offer their work are present. Both user groups need to be present from the beginning to make a successful community. The online marketplace also requires a critical mass of providers and consumers of services to be present at any moment in time. To this end, the marketplace has to make it easy to retrieve and access services. Finally, trust needs to be developed so that the usual risks involved in transactions on the online marketplace can be reduced. In order to cultivate trust, consumers and producers of tasks can often review each other's performances.

Another resource almost completely missing from Digital Humanities organizations is analytical expertise for advanced specialized computing tasks. Kaggle (https://www.kaggle.com/) is a crowd-based infrastructure specializing in delivering such skills. It aims to develop models to solve advanced data problems and make data analytics a 'sport'. Those in need of data analytics can pay Kaggle to host a competition with their data.

Kaggle in turn offers connections to existing analytical talent. Anyone who wants to develop a competition with Kaggle needs to submit their data, with Kaggle offering services to help publish the data on its platform in an effective and ethical way. Participants in a competition then use the data sets, trying their models against them. At the end of the competition, the winner is announced and given the prize. 'Kaggle is a way to organize the brainpower of the world's most talented data scientists and make it accessible to organizations of every size', according to Google's Varian (Reuters, 2011).

Kaggle'scrowd-sourcing approach uses the fact that data problems have many possible solutions, while there are also established techniques to find out which models describe the data problem best. But there remains a mismatch, as not a lot of organisations that collect data have the in-house analytical skills to also exploit this data. The situation is thus similar to many humanities organisations, which have collections but lack the skills to gain insights from it. Again, the commercial model of Kaggle would have to be shifted to one based on voluntary contributions to make Kaggle'scrowd-sourcing work in the humanities. Nevertheless, both online marketplaces and online competitions offer new exciting opportunities for humanities and should thus be part of research infrastructures. We are sure that the next couple of years will see new projects develop here that will fill this gap.

**Conclusion**

This chapter has introduced a number of ideas behind major humanities research infrastructure initiatives in Europe. It has started off by investigating how they address existing criticisms through a new innovative model that can be described as a loosely coupled ecosystem of services and activities. Humanities research infrastructures have been very successful and led to high-profile projects in the last decade in Europe. They have opened up completely new funding schemes to the involved communities and followed a specific need for transnational research. This has been necessary because of the highly diverse and heterogeneous research landscape of humanities in Europe.

Humanities research infrastructures have successfully developed new models of integrating and developing services around communities. It has become clear during these developments that the infrastructures need to entail next to cloud components that offer a digital platform for research also crowd work. This corresponds to the specific ways of working with data in the humanities and the challenges of the underlying complexities of data, which go beyond what is currently possible with computational means. We have analyzed in this chapter how this interaction of crowds and clouds works for six successful large-scale European initiatives. A major distinction here was whether the research communities have to deal with data produced during their own research processes when analyzing cultural objects or data that is offered to them in traditional archives.

We finally tried to look into the future and other crowd-sourcing innovations that can enhance digital humanities research practices. Especially promising are crowd-based marketplaces and competitions. Humanities research infrastructures have been so successful because they could answer a particular community need. We will see many more types and structures develop in the next years that will make use of crowd-based innovations.

**References**

Aloia, N., Papatheodorou, C., Gavrilis, D., Debole, F. & Meghini, C. Describing Research Data: A Case Study for Archaeology. On the Move to Meaningful Internet Systems: OTM 2014 Conferences, 2014. Springer, 768-775.

Anderson, S. & Blanke, T. 2012. Taking the long view: from e-science humanities to humanities digital ecosystems. *Historical Social Research/Historische Sozialforschung***,** 147-164.

Anderson, S., Blanke, T. & Dunn, S. 2010. Methodological commons: arts and humanities e-Science fundamentals. *Philosophical Transactions of the Royal Society A: Mathematical, Physical and Engineering Sciences,* 368**,** 3779-3796.

Aradau, C. & Blanke, T. 2013. The politics of digital crowds. *Lo Squaderno,* 33**,** 31-38.

Basecamp. 2008. *The NO!SPEC campaign vs. crowdSPRING* [Online]. Available: (https://signalvnoise.com/posts/1253-the-nospec-campaign-vs-crowdspring.

Bernipe, P. 2014. *European's Cultural Heritage Online* [Online]. Available: http://epthinktank.eu/2014/04/09/europes-cultural-heritage-online/.

Blanke, T. 2014. *Digital Asset Ecosystems: Rethinking crowds and cloud*, Elsevier.

Blanke, T., Bodard, G., Bryant, M., Dunn, S., Hedges, M., Jackson, M. & Scott, D. Linked data for humanities research—The SPQR experiment. Digital Ecosystems Technologies (DEST), 2012 6th IEEE International Conference on, 2012. IEEE, 1-6.

Blanke, T., Bryant, M. & Hedges, M. Back to our data—Experiments with NoSQL technologies in the Humanities. Big Data, 2013 IEEE International Conference on, 2013. IEEE, 17-20.

Blanke, T., Bryant, M., Hedges, M., Aschenbrenner, A. & Priddy, M. Preparing DARIAH. E-Science (e-Science), 2011 IEEE 7th International Conference on, 2011. IEEE, 158-165.

Blanke, T. & Kristel, C. 2013. Integrating Holocaust Research. *International Journal of Humanities and Arts Computing,* 7**,** 41-57.
Briscoe, G. & Sadedin, S. 2009. Digital Business Ecosystems: Natural science paradigms. *arXiv preprint arXiv:0910.0646*.
Buneman, P., Cheney, J., Tan, W.-C. & Vansummeren, S. Curated databases. Proceedings of the twenty-seventh ACM SIGMOD-SIGACT-SIGART symposium on Principles of database systems, 2008. ACM, 1-12.
Commission, E. 2014. *Research Infrastructures Programme* [Online]. Available: http://ec.europa.eu/research/participants/portal/desktop/en/opportunities/h2020/topics/64-infraia-1-2014-2015.html.
Dombrowski, Q. 2014. What Ever Happened to Project Bamboo? *Literary and Linguistic Computing,* 29**,** 326-339.
Edwards, P.N., Bowker, G.C., Jackson, S.J. & Williams, R. 2009. Introduction: an agenda for infrastructure studies. *Journal of the Association for Information Systems,* 10**,** 6.
Esfri. 2010. *ESFRI Roadmap* [Online]. Available: https://ec.europa.eu/research/infrastructures/pdf/esfri-strategy_report_and_roadmap.pdf.
Gartner, R. & Hedges, M. CENDARI: establishing a digital ecosystem for historical research. Digital Ecosystems and Technologies (DEST), 2013 7th IEEE International Conference on, 2013. IEEE, 61-65.
Holley, R. 2010. Crowdsourcing: how and why should libraries do it? *D-Lib Magazine* [Online], 16. Available: http://www.dlib.org/dlib/march10/holley/03holley.html.
Hyde, A. 2008. *Spec Work Is Evil / Why I Hate CrowdSpring* [Online]. Available: http://andrewhy.de/spec-work-is-evil-why-i-hate-crowdspring/.
Ieee Digital Ecosystem. 2007. *Digital Ecosystem* [Online]. Available: http://www.ieee-dest.curtin.edu.au/2007/ [Accessed 2/1/ 2013].
Kitchin, R. 2014. *The data revolution: Big data, open data, data infrastructures and their consequences*, Sage.
Laumeister, G. 2014. *Why Online Marketplaces Are Booming* [Online]. Available: http://www.forbes.com/sites/groupthink/2014/08/20/why-online-marketplaces-are-booming/.
Malone, T., Laubacher, R. & Dellarocas, C. 2009. Harnessing Crowds: Mapping the Genome of Collective Intelligence. *Working Paper Series* [Online]. Available from: http://cci.mit.edu/publications/CCIwp2009-01.pdf.
Mayer-Schönberger, V. & Cukier, K. 2013. *Big Data: A Revolution that Will Transform how We Live, Work, and Think,* Boston, Houghton Mifflin Harcourt.
Preston, A. 2015. *The war against humanities at Britain's universities* [Online]. Available: http://www.theguardian.com/education/2015/mar/29/war-against-humanities-at-britains-universities.
Reisz, M. 2014. *Humanities research 'needs firm foundations'* [Online]. Available: https://http://www.timeshighereducation.com/news/humanities-research-needs-firm-foundations/2014810.article.
Reuters. 2011. *Kaggle Raises $11 Million in Series A Financing Led by Index Ventures and Khosla Ventures* [Online]. Available: http://www.reuters.com/article/idUS58636+03-Nov-2011+BW20111103.
Richards, J.D. 2012. Digital Infrastructures for Archaeological Research: A European Perspective. *CSA Newsletter* XXV.
Ruppert, E., Law, J. & Savage, M. 2013. Reassembling social science methods: the challenge of digital devices. *Theory, culture & society,* 30**,** 22-46.
Speck, R., Blanke, T., Kristel, C., Frankl, M., Rodriguez, K. & Daelen, V.V. 2014. The Past and the Future of Holocaust Research: From Disparate Sources to an Integrated European Holocaust Research Infrastructure. *arXiv preprint arXiv:1405.2407*.
Svensson, P. 2011. From optical fiber to conceptual cyberinfrastructure. *Digital Humanities Quarterly,* 5.

Thaller, M. 2012. Controversies around the Digital Humanities: An Agenda. *Historical Social Research/Historische Sozialforschung***,** 7-23.
Van Zundert, J. 2012. If you build it, will we come? Large scale digital infrastructures as a dead end for digital humanities. *Historical Social Research/Historische Sozialforschung***,** 165-186.
Váradi, T., Wynne, M. & Koskenniemi, K. CLARIN: Common Language Resources and Technology Infrastructure. 2008.
Zittrain, J. 2008. Ubiquitous human computing. *Philosophical Transactions of the Royal Society A: Mathematical, Physical and Engineering Sciences,* 366**,** 3813-3821.